\documentclass[manuscript]{aastex}
\usepackage{lscape,graphicx,rotating}
\usepackage{geometry}                
\usepackage{amssymb}
\usepackage{amsmath}
\usepackage{natbib}
\shorttitle{Modes of Star Formation in Finite Molecular Clouds}
\shortauthors{Pon et al.}

\begin{document}

\title{Modes of Star Formation in Finite Molecular Clouds}

\author{A. Pon\altaffilmark{1,2}, D. Johnstone\altaffilmark{2,1}, F. Heitsch\altaffilmark{3}}
 
\altaffiltext{1}{Department of Physics and Astronomy, University of Victoria, PO Box 3055 STN CSC, Victoria BC V8W 3P6, Canada; arpon@uvic.ca}
\altaffiltext{2}{NRC-Herzberg Institute of Astrophysics, 5071 W. Saanich Road, Victoria, BC, V9E 2E7, Canada; Douglas.Johnstone@nrc-cnrc.gc.ca}
\altaffiltext{3}{Department of Physics and Astronomy, University of North Carolina Chapel Hill, CB 3255, Phillips Hall, Chapel Hill, NC 27599, USA; fheitsch@unc.edu}

\begin{abstract}
We analytically investigate the modes of gravity-induced star formation possible in idealized finite molecular clouds where global collapse competes against both local Jeans instabilities and discontinuous edge instabilities. We examine these timescales for collapse in spheres, discs, and cylinders, with emphasis on the structure, size, and degree of internal perturbations required in order for local collapse to occur before global collapse. We find that internal, local collapse is more effective for the lower dimensional objects. Spheres and discs, if unsupported against global collapse, must either contain strong perturbations or must be unrealistically large in order for small density perturbations to collapse significantly faster than the entire cloud. We find, on the other hand, that filamentary geometry is the most favorable situation for the smallest perturbations to grow before global collapse overwhelms them and that filaments containing only a few Jeans masses and weak density perturbations can readily fragment. These idealized solutions are compared with simulations of star-forming regions in an attempt to delineate the role of global, local, and edge instabilities in determining the fragmentation properties of molecular clouds. The combined results are also discussed in the context of recent observations of Galactic molecular clouds.
 \end{abstract}


\section{INTRODUCTION}
\label{intro}

Stars primarily form in clustered environments inside of molecular clouds containing tens of thousands of solar masses of material or more \citep{Lada03Lada}. These star-forming molecular clouds must fragment to give rise to the observed initial mass function (IMF) of stars, with a characteristic mass on the order of a solar mass (e.g., \citealt{Bonnell07}), and cannot be solely dominated by large scale, global collapse modes. Gravitational collapse within a molecular cloud leads to the formation of individual stars; however, the mechanism driving the formation and shape of the IMF is poorly constrained. While turbulence is believed to play a key role in producing significant inhomogeneity from which fragmentation can occur, gravity must also contribute (e.g., \citealt{Bonnell07}). In fact, the idea that gravity is driving supersonic motions (or ``turbulence'') in molecular clouds has been around for awhile (e.g., \citealt{Goldreich74, Ferrini83, Falgarone86}). Beyond the monolithic collapse envisioned by \citet{Zuckerman74} in their critique of \citet{Goldreich74}, recent numerical and analytical studies show that global gravity can indeed drive ``turbulent'' motions without leading directly to ``catastrophic'' collapse (e.g., \citealt{Hartmann01, Hartmann02, BallesterosParedes07, VazquezSemadeni07, Field08, Heitsch08}). Depending on the morphology of the cloud, the fragmentation that occurs during such a large scale collapse might also lead to an IMF distribution of stars.  In this scenario, the observed supersonic motions of the gas are determined by the geometry of the cloud, which depends strongly upon the initial conditions of the cloud.

Spatially resolved observations of star-forming molecular clouds usually exhibit rather complex geometries, including substructure often associated with lower dimensionality, such as sheets and filaments (e.g., \citealt{Schneider79, Bally87, Lada99, Hartmann02, Johnstone03, Lada07, Myers09, Molinari10, Andre10}). Filamentary structures have also been noted to form in molecular cloud simulations, although their formation mechanism varies with the type of model used, be it driven or decaying supersonic turbulence (e.g., \citealt{Padoan07, Bate09}), flow collisions (e.g., \citealt{VazquezSemadeni10}), or global gravitational accelerations (e.g., \citealt{Burkert04, Hartmann07}). 

In the absence of turbulent motions, the collapse properties of isothermal spheres (e.g., \citealt{Larson69, Penston69II, Shu77}), infinitely large, isothermal sheets (e.g., \citealt{Larson85, Myers09}), and infinitely long, isothermal cylinders (e.g., \citealt{Stodolkiewicz63, Ostriker64, Inutsuka92, Inutsuka97, Nakamura93, Fiege00}) have been thoroughly studied. Equilibrium configurations of infinite sheets perturbed to form successive, parallel filaments and infinite, isothermal filaments perturbed to form strings of dense cores have also been found \citep{SchmidBurgk67, Curry00, Myers09}. Yet, the collapse of {\em finitely} long cylinders and discs, and the subsequent interplay between global and local collapse modes, are much less well understood. Previous simulations of finite cylinders show that these cylinders can fragment into multiple condensations (e.g., \citealt{Bastien83, Bastien91}), while the disc simulations of \citet{Burkert04} show no sign of local collapse modes. These disc simulations are dominated by the global collapse of the disc and the subsequent build up of mass along the outer ring of the disc (see also \citealt{Hsu10}).

In this paper, we compare analytic estimates for the global collapse timescales to the local collapse timescales of small density perturbations embedded in different dimensional objects. We analyze whether small density perturbations are capable of collapsing significantly faster than the entire cloud for spherical, disc shaped, and filamentary clouds to determine whether strong perturbations, possibly induced by turbulent motions, are required to fragment these differently shaped molecular clouds. We present our analytic results in \S \ref{analytic}. In \S \ref{discussion}, we discuss these results in context of recent observations and simulations and we present a summary of our findings in \S \ref{conclusion}.

\section{MODES OF COLLAPSE - ANALYTIC THEORY}
\label{analytic}

To understand the importance of global gravity in molecular clouds, we consider the large scale gravitational forces at play in comparison to the growth of local perturbations within a cloud. For a uniform sphere of arbitrary radius, the collapse is quickest for the largest scale (e.g., \citealt{Larson85}), since gravity is a long-range force and global collapse is inevitable provided the region contains more than a Jeans mass of material and support from effects such as magnetic fields and rotation are negligible. In two-dimensions, however, an infinite sheet can be constructed for which there is no global gravitational force in the plane and the collapse is dominated by objects with radii similar to the thickness of the sheet \citep{Larson85}. A similar argument can be made for infinite one-dimensional cylinders \citep{Larson85, Inutsuka92, Inutsuka97}. As noted by \citet{Burkert04} the lack of global gravity in these systems depends entirely on the infinite nature of the sheet or cylinder. In the following we show how the global gravitational collapse depends on both the dimensionality and the finite nature of simple geometries: spheres, discs, and cylinders.

\subsection{Thermal Support}
\label{thermal}

Thermal support is always present in a molecular cloud and will oppose gravitational collapse. This thermal support is most effective on small scales and can be overcome on larger scales. The Jeans length is the size scale above which, in three dimensions, thermal pressure is no longer capable of supporting an isothermal cloud. The Jeans length, $R_J$, is given by:
\begin{equation}
R_J = c_s \left(\frac{\pi}{G \, \rho}\right)^{\frac{1}{2}},
\end{equation}
where $c_s$ is the isothermal sound speed, G is the gravitational constant, and $\rho$ is the mass density of the gas. For isothermal, infinite sheets, a stable equilibrium configuration can be found for all surface densities such that thermal pressure counteracts gravity. In these equilibrium configurations, the scale height of the gas is on the order of the Jeans length of the material in the midplane \citep{Spitzer42, Spitzer78} and small perturbations in the plane of the disc will only collapse if they are larger than the scale height of the gas \citep{Larson85}. For infinite, isothermal cylinders, the characteristic radius of an equilibrium configuration and the size scale on which thermal support is effective are on the order of the Jeans length at the central axis of the cylinder \citep{Stodolkiewicz63, Ostriker64}. Unlike infinite isothermal sheets, infinite isothermal cylinders have a critical mass per unit length value above which no equilibrium configuration exists \citep{Ostriker64}. 

In the following sections, we only consider the effect of gravity and do not explicitly include any thermal effects in calculating collapse timescales, although we do interpret our results in light of the Jeans length being the minimum scale which can collapse. We also do not consider the effects of turbulence, rotation, magnetic fields, or deviations from isothermality.

\subsection{3D - Sphere}
\label{3d}

For a sphere of density $\rho$ and radius $R \gg R_J$, the instantaneous global collapse time, $t_g$, is given by
\begin{eqnarray}
t_g &=&  \left( \frac{3}{2\pi \,G \, \rho} \right)^{1/2}.
\end{eqnarray}
The above instantaneous global collapse time differs by less than a factor of two from the collapse timescale obtained by properly following the changes in the strength of the gravitational acceleration over time (e.g., \citealt{Carroll07}). A small region, $R_1 \ll R$, at the centre of the sphere with an enhanced density $(1+\epsilon)\rho$, where $\epsilon \ll 1$, might act as a seed for local collapse. Since the density is only slightly enhanced, the Jeans length for this region is very similar to the unperturbed Jeans length and the region is stable against {\it local} collapse when $R_1 < R_J$. However, provided $R_1 > R_J$, the local collapse time, $t_l$, for this inner region approaches
\begin{eqnarray}
t_l &=&  \left( \frac{3}{2\pi \,G \, (1+\epsilon)\rho} \right)^{1/2},
\end{eqnarray}
where we have ignored the residual support due to thermal pressure. Comparing these two timescales reveals the difficulty in fragmenting an almost uniform spherical molecular cloud:
\begin{eqnarray}
t_l = (1+\epsilon)^{-1/2}\,t_g.
\label{eqn:tl3d}
\end{eqnarray}
For small amplitude density perturbations, there is little time for the local collapse to proceed before the global collapse consumes the entire spherical cloud. In three dimensions, local collapse beats global collapse {\it only} if there are significant  inhomogeneities (strong perturbations) in the medium. \citet{Tohline80b} showed that the growth of this perturbation is even slower, relative to the global collapse timescale, if the perturbation is induced after global collapse has begun.
	
\subsection{2D - Disc}
\label{2d}

\subsubsection{Accelerations}
\label{2da}

For an infinitely thin disc with constant surface density $\Sigma$ and radius $R \gg R_J$, the central acceleration at a distance of $r  > R$ from the center of this disc is
 \begin{equation}
\label{eqn:ae1}
a_{\rm ext} = 4G \, \Sigma \left[K\left(\frac{R} {r}\right) - E\left(\frac{R}{r}\right)\right],
\end{equation}
where  K and E are the first and second complete elliptic integrals respectively \citep{Wyse42}. The central acceleration of a point inside of the disc, i.e., $r < R$, is \citep{Wyse42}:
\begin{equation}
\label{eqn:ae2}
a_{\rm int} = 4G \, \Sigma \frac{R}{r} \left[K\left(\frac{r} {R}\right) - E\left(\frac{r}{R}\right)\right].
\end{equation}

For a perturbed disc with surface density  $(1 + \epsilon)\, \Sigma$ inside of $R_1 \ll R$, the net acceleration is
\begin{equation}
\label{eqn:anet2}
a_{\rm net} = 
\begin{cases}
4G \, \Sigma \frac{R}{r} \Bigg( 
\Big[K\big(\frac{r} {R}\big)   - E\big(\frac{r}{R}  \big)\Big] 
+ \epsilon \frac{R_1}{R}\, 
\Big[K\big(\frac{r} {R_1}\big) - E\big(\frac{r}{R_1}\big)\Big]
\Bigg), & r < R_1\\[0.25 in]
4G \, \Sigma \frac{R}{r} \Bigg( 
\Big[K\big(\frac{r} {R}\big)   - E\big(\frac{r}{R}  \big)\Big] 
+ \epsilon \frac{r}{R}\, 
\Big[K\big(\frac{R_1} {r}\big) - E\big(\frac{R_1}{r}\big)\Big]
\Bigg), & R_1 < r < R.
\end{cases}
\end{equation}
For the case of $R_1 < r < R$, \citet{Burkert04} showed that these elliptic integrals can be expanded to lowest order to give:
\begin{equation}
\label{eqn:aa}
a_{\rm net} =  \frac{\pi G \, \Sigma \, r}{R} \left(1 + \frac{\epsilon R R_1^2} {r^3}\right).
\end{equation}

\begin{figure}[htbp] 
   \centering
   \includegraphics[width=5.5in]{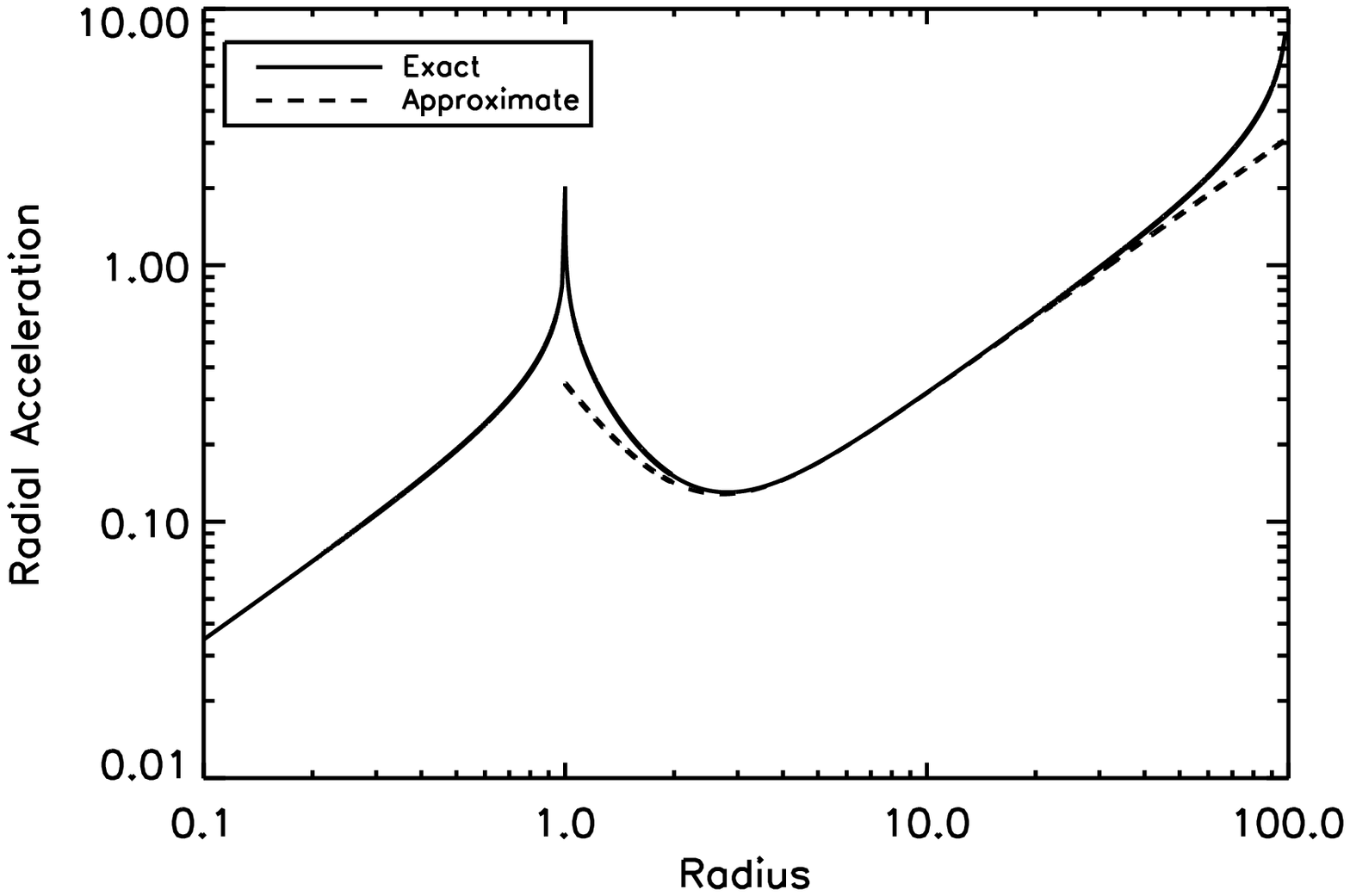}
   \caption{Accelerations of a perturbed disc with $R=100$, $G=1$, $\Sigma=1$, $\epsilon=0.1$, and $R_1=1$. The solid line shows the exact solution for the central accelerations while the dashed line shows the lowest order approximation to the exact solution.}
   \label{fig:2da}
\end{figure}

Figure \ref{fig:2da} compares the accelerations of a perturbed disc calculated from the above exact and approximate solutions. For simplicity, $G$, $\Sigma$, and $R_1$ have all been set to 1, $\epsilon$ has been set to 0.1, and $R$ has been set to 100. The approximate solution of Equation \ref{eqn:aa} reproduces the exact accelerations of Equation \ref{eqn:anet2} for intermediate radii between $R_1$ and $R$, but underestimates the accelerations closer to $R_1$ and $R$. 

\subsubsection{Timescales}
\label{2dt}

The square of the instantaneous collapse time at $r$ is
\begin{eqnarray}
\label{eqn:t}
t^2(r) &=& \frac{2r}{a_{net}(r)},\notag\\[0.15 in]
\label{eqnt2r}
t^2(r) &=& \frac{2R}{\pi G \, \Sigma} \left(1 + \frac{\epsilon \, R \, R_1^2}{r^3}\right)^{-1}.
\end{eqnarray}

The global collapse timescale, $t_g$, can be determined by considering the collapse time when $r = R$. That is
\begin{eqnarray}
t^2_g(R) &=&  \frac{2R}{\pi G \, \Sigma} \left(1 + \frac{\epsilon \, R_1^2}{R^2}\right)^{-1}.
\end{eqnarray}
Given that the initial surface density perturbation is small, $\epsilon \ll 1$, 
and $R_1 \ll R$ :
\begin{eqnarray}
\frac{\epsilon \, R_1^2}{R^2} &\ll& 1,\notag \\[0.15 in]
t^2_g &=&\frac{2R}{\pi G \, \Sigma} .
\end{eqnarray}
Thus, the global timescale is not changed significantly due to the small perturbation. 

Equation \ref{eqnt2r} can now be re-written as
\begin{equation}
\label{eqn:t2}
t^2(r) = t^2_g \left(1 + \frac{\epsilon \, R \, R_1^2}{r^3}\right)^{-1}.
\end{equation}
Note that the collapse time $t(r)$ is equal to the global collapse time and independent of $r$ unless $\epsilon > 0$. When there is a density enhancement, however, the local time for collapse is dependent on $r$ and thus, smaller regions can collapse faster than the global collapse.

Assuming that the perturbed region is large enough to be Jeans unstable, $R_1 > R_J$, the timescale for collapse of a local region with radius $R_1$ is approximately
\begin{eqnarray}
t^2_l &=& t^2(R_1),\notag\\[0.15 in]
t^2_l &=& t^2_g \left(1 + \frac{\epsilon \, R}{R_1}\right)^{-1}.
\label{eqn:2dlocal} 
\end{eqnarray}
For this to be significantly smaller than the global collapse timescale
\begin{eqnarray}
\frac{\epsilon \, R}{R_1} &\gg& 1, \notag\\[0.15 in]
\epsilon &\gg& \frac{R_1}{R}.
\end{eqnarray}
Thus, it is possible to find a combination of perturbation length and amplitude that results in significantly enhanced local collapse. For a small density perturbation, $\epsilon = 0.1$, the perturbation length must be approximately 100 times smaller than the disc's total radius in order for the local collapse timescale to be 3 times smaller than the global collapse timescale.  Recognizing that thermal support in the
disc requires $R_1 > R_J$ for local collapse to occur, such a disc would need to have assembled more than $10^4$ Jeans masses of material! Even for a large perturbation, $\epsilon = 1$, the disc would still need more than 100 Jeans masses for the local collapse mode to be 3 times faster than the global collapse mode.

\begin{figure}[htbp] 
   \centering
   \includegraphics[width=5.5in]{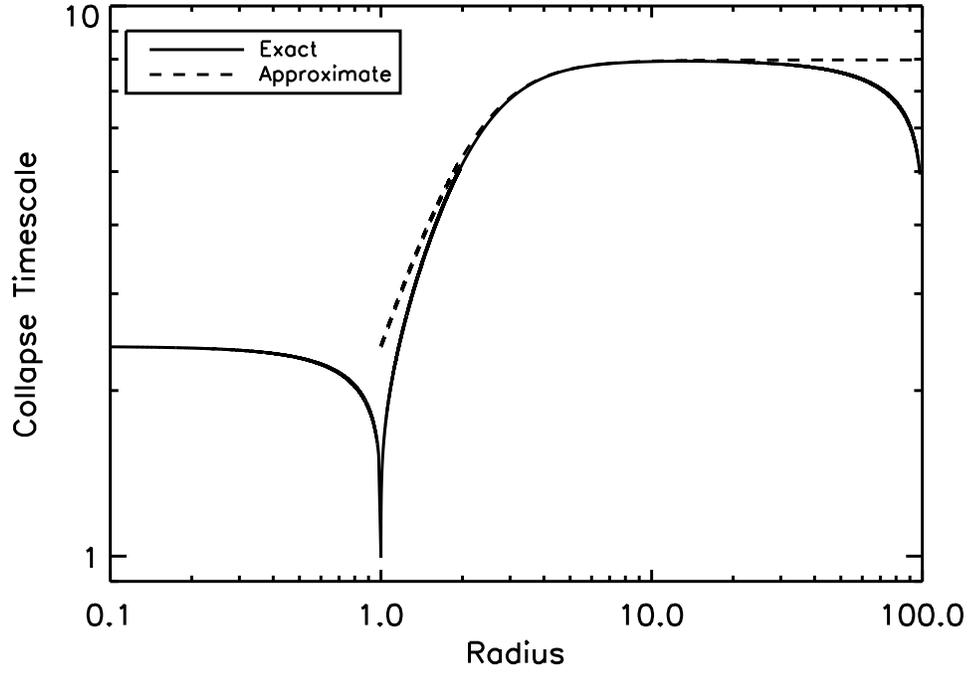}
   \caption{Collapse timescales of a perturbed disc with $R=100$, $G=1$, $\Sigma=1$, $\epsilon=0.1$, and $R_1=1$. The solid line shows the timescales calculated from the exact accelerations while the dashed line shows the timescales calculated from the approximate accelerations.}
   \label{fig:2dt}
\end{figure}

Figure \ref{fig:2dt} shows the collapse timescales for a disc with $R_1 = G = \Sigma = 1$, $\epsilon = 0.1$, and $R/R_1 = 100$. The solid line shows the timescales calculated from the exact accelerations while the dashed line shows the timescales calculated from the approximate accelerations given by Equation \ref{eqn:aa}. The timescales for collapse interior to a radius of $R_1$ are approximately a factor of three faster than the collapse timescales for the outer parts of the disc and the timescale for collapse at $R_1$, as calculated from the approximate accelerations, is roughly equal to the collapse timescales for most points inside of $R_1$ as calculated from the exact accelerations.

The exact accelerations lead to collapse timescales approaching zero at radii of $R$ and $R_1$, while the approximate accelerations produce timescales that do not show this asymptotic behavior. These extremely small collapse times are artifacts of having an infinitely thin disc with infinitely sharp edges and no thermal support. Real discs have finite heights and smoother edges, which should suppress this asymptotic behavior. We will come back to the issue of these infinitely small collapse timescales at the edges in Section \ref{edge}.

Thermal pressure provides for support against collapse on scales smaller than the Jeans length, $R_J$, and thus, the above analysis suggests that the preferred size of the perturbed region is $R_1 \sim R_J$, since this provides the greatest leverage on the local to global collapse timescale.  This length scale is also associated with the scale height of a thermally supported disc and is close to the preferred scale for disc fragmentation in infinite sheets \citep{Larson85}.

\subsection{1D - Cylinder}
\label{1d}

\subsubsection{Accelerations}
\label{1da}

	For a one dimensional cylinder of length $2L$ and line density $\lambda$, the net acceleration a distance $d < L$ from the center of the cylinder, along the central axis of the cylinder, is
\begin{eqnarray}
a_{int}(d) &=& \int_0^{L+d} \frac{G \lambda \, dr} {r^2} - \int_0^{L-d} \frac{G \lambda \, dr} {r^2} , \notag\\[0.15 in]
a_{int}(d) &=& \int_{L-d}^{L+d} \frac{G \lambda \, dr} {r^2}, \notag\\[0.15 in]
a_{int}(d) &=& \frac{G \lambda}{L - d} - \frac{G \lambda} {L + d}.
\end{eqnarray}

The net acceleration a distance $d > L$ from the center of the cylinder, and along the central axis of the cylinder, is
\begin{eqnarray}
a_{ext}(d) &=& \int_{d-L}^{L+d} \frac{G \lambda \, dr} {r^2}, \notag\\[0.15 in]
a_{ext}(d) &=& \frac{G \lambda} {d - L} - \frac{G \lambda} {L + d}.
\end{eqnarray}

	If the above described cylinder is perturbed such that its density becomes $(1 + \epsilon)\, \lambda$ within a distance of $L_1$ of the center of the cylinder, then the acceleration a distance $d$ from the center will be
\begin{equation}
a_{\rm net}(d) = 
\begin{cases}
\frac{G \lambda}{L - d} - \frac{G \lambda} {L + d} + \frac{G \lambda \, \epsilon}{L_1 - d} - \frac{G \lambda \, \epsilon} {L_1 + d}, & d < L_1\\[0.15 in]
\frac{G \lambda}{L - d} - \frac{G \lambda} {L + d} + \frac{G \lambda \, \epsilon} {d - L_1} - \frac{G \lambda \epsilon} {L_1 + d}, & L_1 < d < L.
\end{cases} 
\end{equation}

For the case of $L_1 < d < L$, expanding to lowest order gives
\begin{eqnarray}
a_{\rm net}(d) &=& G \lambda \left(\frac{1}{L} \left[(1 + \frac{d}{L}) - (1-\frac{d}{L}) \right]+\frac{\epsilon}{d}\left[(1+ \frac{L_1}{d}) - (1-\frac{L_1}{d})\right]\right),\notag\\[0.15in]
a_{\rm net}(d) &=& \frac{G \lambda \, 2d}{L^2} \left(1+\frac{\epsilon\,L_1\,L^2}{d^3}\right).
\label{eqn:1dapproxa}
\end{eqnarray}

\begin{figure}[htbp] 
   \centering
   \includegraphics[width=5.5in]{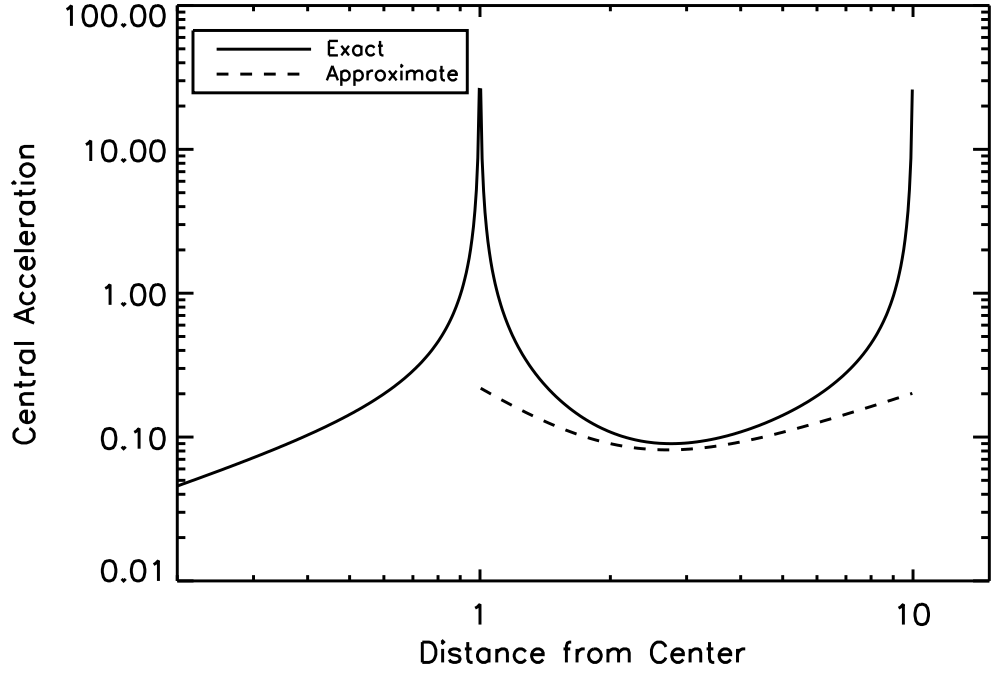}
   \caption{Accelerations of a perturbed cylinder with $L_1=1$, $G=1$, $\lambda=1$, $\epsilon=0.1$, and $L=10$. The solid line shows the exact solution for the central accelerations while the dashed line shows the lowest order approximation to the exact solution.}
   \label{fig:1da}
\end{figure}

Figure \ref{fig:1da} shows the accelerations of a perturbed cylinder as calculated from the above exact and approximate equations. For simplicity, $\lambda$, $G$, and $L_1$ are set to 1 while $\epsilon$ and $L $ are set to 0.1 and 10 respectively. Just as with the disc, Equation \ref{eqn:1dapproxa} accurately reproduces the accelerations at intermediate radii but underestimates the accelerations near $L_1$ and $L$.

\subsubsection{Timescales}
\label{1dt}

The square of the instantaneous collapse time is approximately
\begin{eqnarray}
t^2(d) &=& \frac{L^2}{G \lambda} \left(1+\frac{\epsilon\,L_1\,L^2}{d^3}\right)^{-1}.
\label{eqn:t1d}
\end{eqnarray}
The square of the global collapse time is thus
\begin{eqnarray}
t^2_g&=&t^2(L),\notag\\[0.15in]
t^2_g &=& \frac{L^2}{G \lambda} \left(1+\frac{\epsilon\,L_1\,}{L}\right)^{-1}.
\end{eqnarray}
For small perturbations
\begin{eqnarray}
\frac{\epsilon\,L_1\,}{L} &\ll& 1,\notag\\[0.15 in]
t^2_g &=& \frac{L^2}{G \lambda}.
\label{eqn:globalfil}
\end{eqnarray}
Just as with the disc, the global collapse time is essentially the same in a perturbed and unperturbed cylinder and Equation \ref{eqn:t1d} can be written as
\begin{equation}
t^2(d) = t^2_g \left(1+\frac{\epsilon\,L_1\,L^2}{d^3}\right)^{-1}.
\end{equation}
When $\epsilon = 0$, the local collapse time is essentially the global collapse time. The presence of a line density perturbation speeds up the local internal collapse. 

 The square of the collapse timescale for a local region of size $L_1$ is
\begin{eqnarray}
t^2_l&=&t^2(L_1),\notag\\[0.15in]
t^2_l &=&t_g^2 \left(1+\frac{\epsilon L^2}{L_1^2}\right)^{-1}.
\label{eqn:1dlocal}
\end{eqnarray}
Equation \ref{eqn:1dlocal} indicates that the smaller the perturbed region is in comparison to the total length of the cylinder, the faster the local collapse will be in comparison to global collapse.

For the local collapse timescale to be significantly shorter than the global collapse timescale
\begin{eqnarray}
\frac{\epsilon L^2}{L_1^2} &\gg& 1,\notag\\[0.15in]
L_1 &\ll& \sqrt{\epsilon} L.
\end{eqnarray}
More specifically, for the local collapse timescale to be a factor of three smaller than the global collapse timescale,
\begin{eqnarray}
\left(\frac{t_g}{t_l}\right)^2 &>& 10,\notag\\[0.15 in]
L_1 &\lesssim& 0.3\sqrt{\epsilon} L.
\end{eqnarray}

If $\epsilon=0.1$, then the perturbation length must be approximately 10 times smaller than the cylinder's length in order for the local collapse timescale to be three times smaller than the global collapse timescale. Taking $L_1 = R_J$ as the minimum size for a gravitationally unstable weak perturbation, this implies a minimum finite cylinder mass of a few tens of Jeans masses. A large perturbation, $\epsilon = 1$, would require a cylinder with approximately ten Jeans masses in order for the local collapse timescale to be three times smaller than the global collapse timescale.

%the radius should be rJ ~ cs * (1 / grho)^0.5? With R = Rj, and MJ = rJ^2, I get MJ = 56 and 18 -> if L_1 = RJ /2, divide by 2. If R = rJ / 2, divide by 4, if MJ defined from sphere, multiply by 2.

%

\begin{figure}[htbp] 
   \centering
   \includegraphics[width=5.5in]{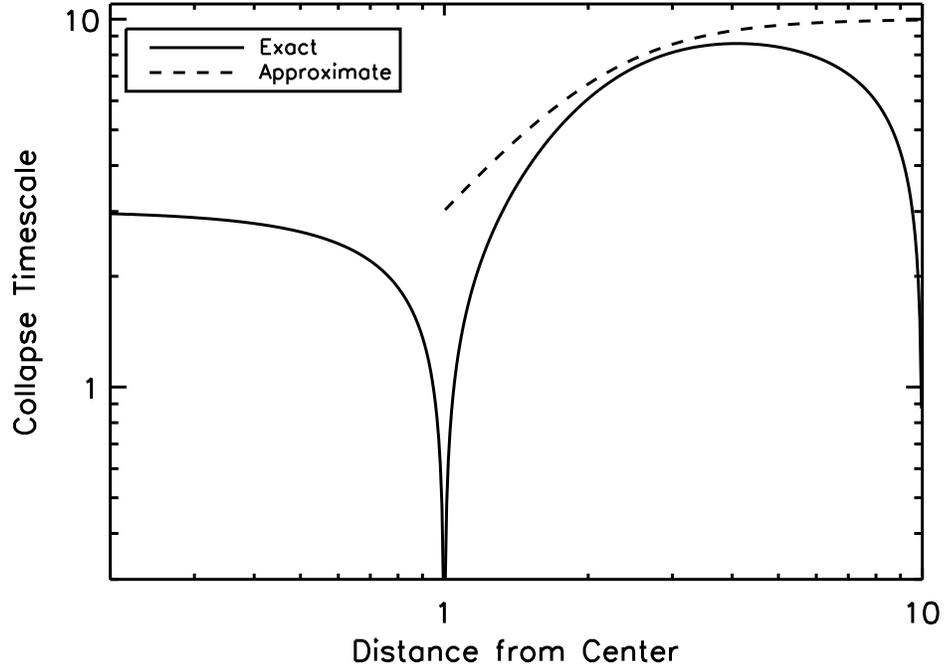}
   \caption{Collapse timescales of a perturbed cylinder with $L_1=1$, $G=1$, $\lambda=1$, $\epsilon=0.1$, and $L=10$. The solid line shows the timescales calculated from the exact accelerations while the dashed line shows the timescales calculated from the approximate accelerations.}
   \label{fig:1dt}
\end{figure}

Figure \ref{fig:1dt} shows the collapse timescales for a perturbed cylinder with $L_1 = G = \lambda = 1$, $\epsilon = 0.1$, and $L/L_1 = 10$. The solid line shows the timescales calculated from the exact accelerations while the dashed line shows the timescales calculated from the approximate accelerations given by Equation \ref{eqn:1dapproxa}. The collapse times for points interior to $L_1$ are approximately a factor of three smaller than the collapse times for most points beyond $L_1$. Just as for the disc, the collapse timescale abruptly drops to zero at $L$ and $L_1$ due to the sharp jumps in density at these points. In more realistic filaments, these density changes would be more gradual and the actual collapse timescales would more closely resemble the collapse timescales derived from the approximate accelerations of Equation \ref{eqn:1dapproxa}. As in the 2D case, the collapse timescale at $L_1$ calculated from the approximate accelerations is roughly equal to the collapse timescales for most points interior to $L_1$, as calculated from the exact accelerations.

\subsection{Edge Effects - A Hybrid Collapse Mode}
\label{edge}

	In the above models, the collapse timescales become infinitely small at density boundaries. This behavior is clearly unphysical and originates from our assumption that the disc and filament are infinitely thin. To investigate the significance of this assumption with respect to our results, we have calculated the accelerations along the central axis of a cylinder with uniform density $\rho$, total height $2L$, and radius $R$. We find that the acceleration a distance $d < L$ from the center of the cylinder, but still along the central axis, is
\begin{equation}
a(d) = 2 \pi \, G \, \rho \, \left(2d + \sqrt{R^2 + (L- d)^2} - \sqrt{R^2+(L+d)^2}\right).
\end{equation}
Evaluating this at the edge of the cylinder, $d = L$, gives a {\it finite} acceleration of
\begin{equation}
a(d) = 2 \pi \, G \, \rho \, \left(2L + R - \sqrt{R^2+4L^2}\right),
\end{equation}
thus proving that it was the infinite thinness of our prior models that led to the infinitely small collapse timescales.

\begin{figure}[htbp] 
   \centering
   \includegraphics[width=5.5in]{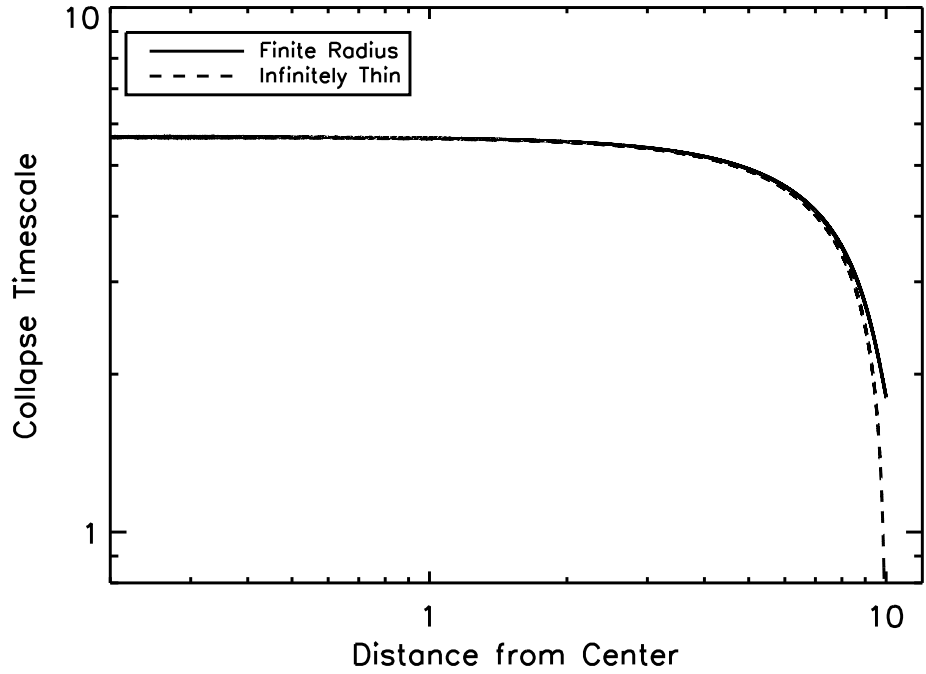}
   \caption{Collapse timescales of a cylinder with $L=10$, $G=1$, $\rho=1$, and $R=1$ are shown as the solid line and the collapse timescales of an infinitely thin cylinder with $L=10$, $G=1$, and $\lambda=\pi$ are shown as the dashed line.}
   \label{fig:comparefiltocyl}
\end{figure}

	As before, these accelerations can easily be turned into instantaneous collapse timescales. Figure \ref{fig:comparefiltocyl} shows the collapse timescales for two cylinders, both with $G = 1$ and $L = 10$, where one cylinder has a radius $R = 1$ and uniform volume density $\rho = 1$ and the other cylinder is infinitely thin and has a line density equivalent to that of the other cylinder, namely $\lambda = \pi$. The collapse timescales of the two cylinders are essentially identical except for points very close to the end of the cylinders. The collapse timescales of the finite cylinder only exceed those of the infinitely thin cylinder by a factor of two or more for points within 0.02 L of the end. Since there is such a good agreement between the collapse timescales and since our global collapse timescales were calculated from our approximate accelerations, which did not trace the sharp increases in acceleration at the edges, our above analysis of the ratio of local to global collapse timescales should be valid for both infinitely thin and finite sized objects. 

	Figure \ref{fig:comparefiltocyl} also shows that while a finite radius cylinder does not have infinite accelerations at its edges, the timescale for collapse still decreases towards the edge in the outer part of the cylinder. For the particular set of parameters of the finite radius cylinder shown in Figure \ref{fig:comparefiltocyl}, the edge collapse timescale is roughly a factor of three less than the global collapse timescale, given by Equation \ref{eqn:globalfil}, for a filament with this line density. This reduction in collapse time at the edge is due to the density contrast at the edge. 
	
	A collapse timescale that decreases outwards leads to material at the edge running into material further inwards, which causes a build up of material. Such a density enhancement at the periphery of a cloud would then collapse on the local collapse timescale. We consider this edge collapse mode to be a third collapse mode that is a combination of a global and local collapse mode, since the initial accelerations are provided by the global gravitational potential while the final collapse only occurs within the local density enhancements along the edge of the cloud.	
	
	The majority of a filamentary cloud should still collapse roughly on the global collapse timescale, rather than the edge collapse timescale, as the material inward of the edge will slow the edge's collapse. This is further supported by the fact that for the cylinders in Figure \ref{fig:comparefiltocyl}, the collapse timescales for over 85\% of the cylinders' lengths are within a factor of two of the global collapse timescale.
	
\begin{figure}[htbp] 
   \centering
\centerline{\resizebox{0.6\hsize}{!}{\includegraphics{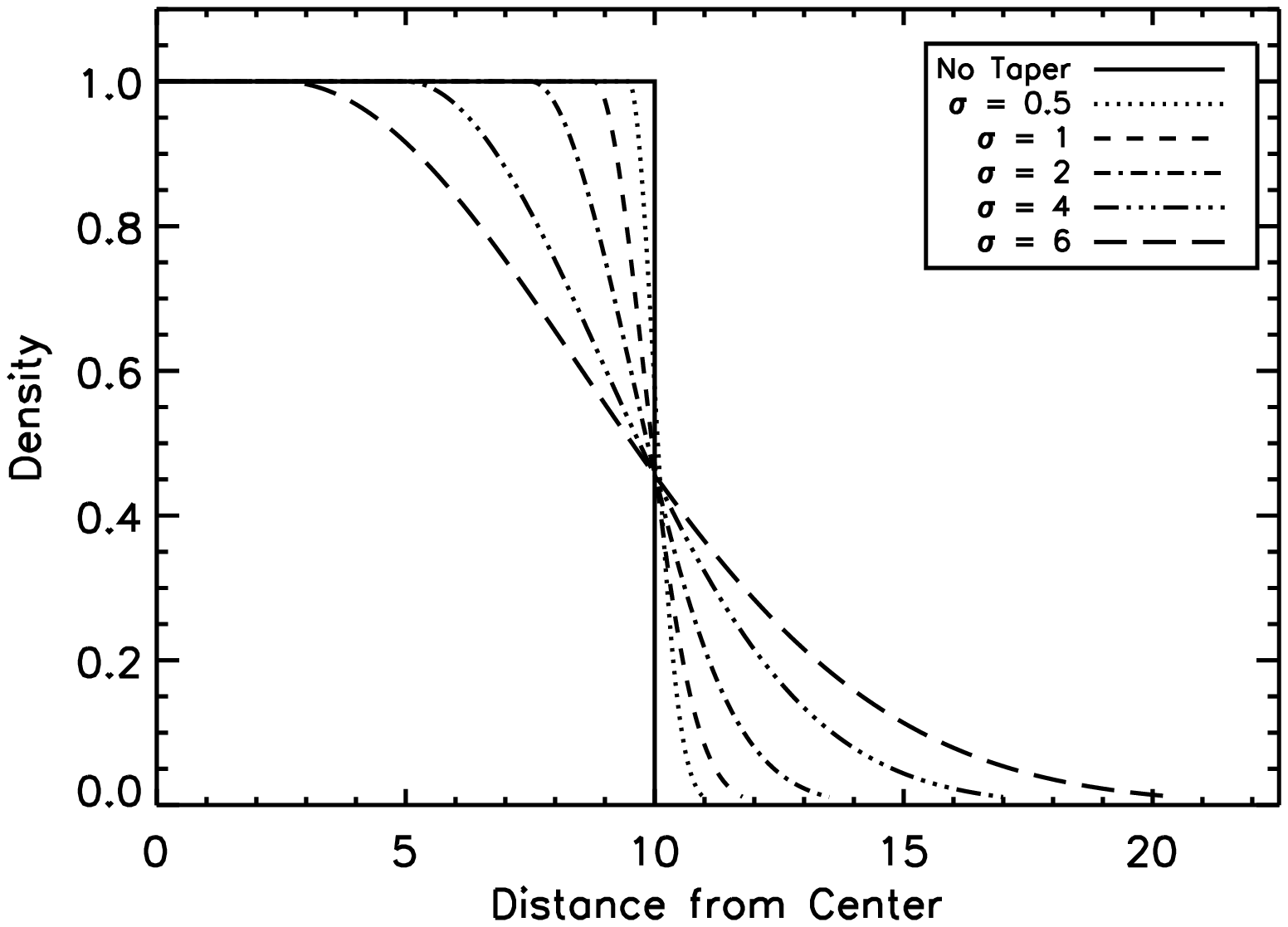}}
            \resizebox{0.6\hsize}{!}{\includegraphics{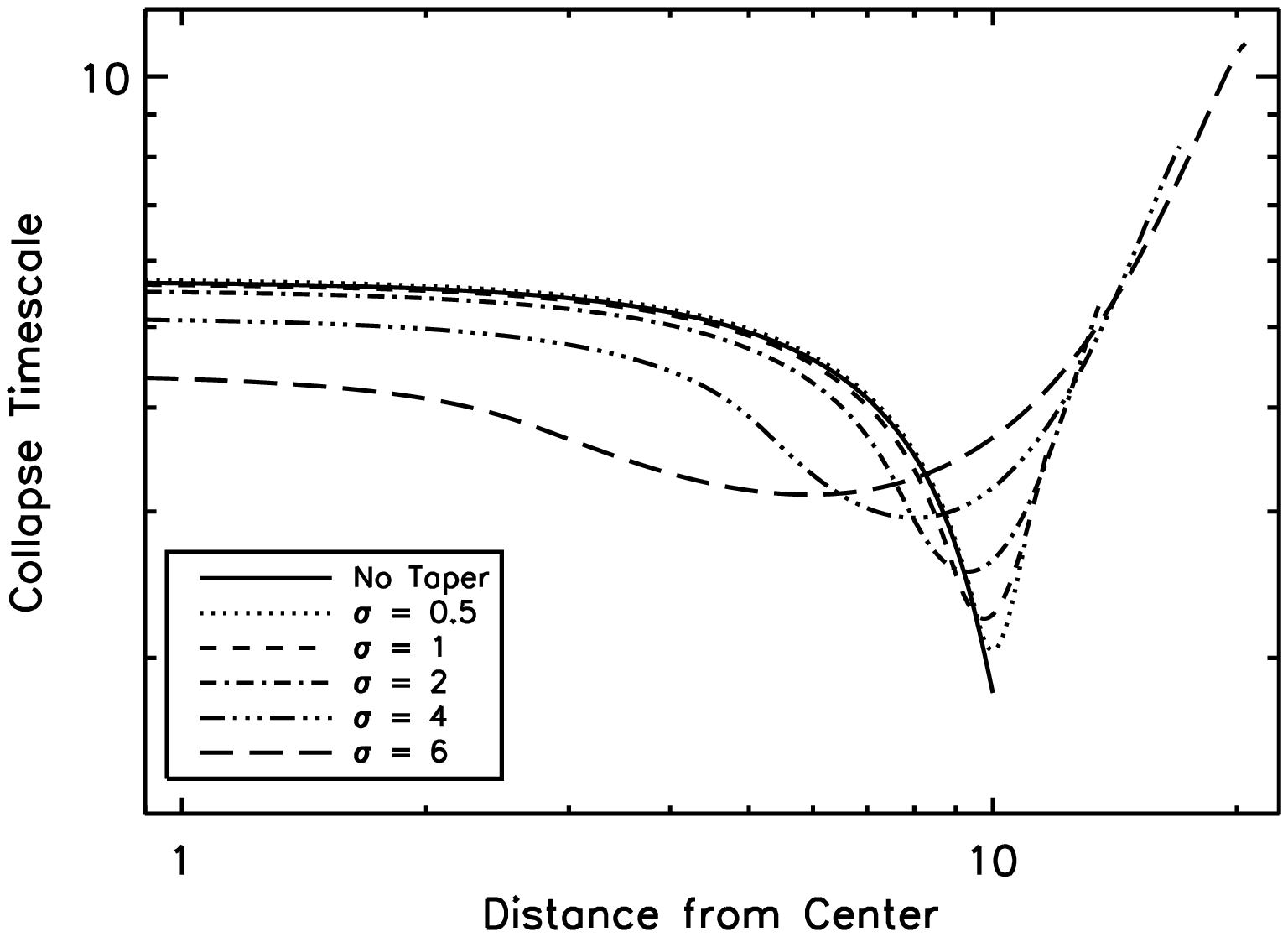}}}     
   \caption{Density profiles (left) and collapse timescales (right) of tapered cylinders. For all cylinders, $G=1$, $R=1$, and the central density is $\rho = 1$. The solid line is for a cylinder with $L=10$ and sharp edges. The dotted, dashed, dot-dash, dash-triple dot, and long dashed lines correspond to cylinders where $\sigma$ is 0.5, 1, 2, 4, and 6, respectively, and where $\sigma$ is the standard deviation of the gaussian profile used for the tapered edge. The location of the tapered edge has been chosen so that all of the cylinders have the same total mass. Note how the collapse timescale still decreases significantly towards the edge, even for the most strongly tapered cylinder.}
   \label{fig:taper}
\end{figure}

	To test how important the sharp edges of our clouds are to the presence of an edge collapse mode, we examine a series of finite radius cylinders in which the density slowly decreases to zero, rather than abruptly dropping to zero, near the faces of each cylinder. We use a cylinder with R = 1, L = 10, $\rho$ = 1, and G = 1 as our comparison, sharp edged cylinder. To create a tapered cylinder, we set the central density of the cylinder, $\rho_0$, to be equal to 1 and outside of a distance $z_{edge}$ from the center of the cylinder, as measured along the central axis of the cylinder, we adopt a gaussian density profile:
\begin{equation}
\rho = \rho_0 \exp\left(\frac{-(z - z_{edge})^2}{2 \sigma^2}\right).
\end{equation}
We examine five tapered cylinders where the values of $\sigma$ are 0.5, 1, 2, 4, and 6. Note that the case of $\sigma = 1$ corresponds to the taper having a size similar to the radius of the cylinder. We extend the tapered edge out to a length of 3$\sigma$, such that the density drops to less than 2\% of the cylinder's central density. We choose the location where the tapered edge starts, $z_{edge}$, by requiring that the total mass of the tapered cylinder is the same as that of our comparison sharp edged cylinder.  We then numerically calculate the instantaneous accelerations along these cylinders and convert these accelerations into collapse timescales, as done previously in this paper. The density profiles and collapse timescales for these cylinders are shown in Figure \ref{fig:taper}.
	
	As the length scale of the taper increases, the collapse timescale towards the edge increases and the collapse timescale of the interior decreases, thereby reducing the effectiveness of the edge collapse mode, as expected. A decrease in collapse timescale towards the edge, however, is still very prominent in all of the tapered cylinders. The ratio of the minimum collapse timescale (edge collapse) to the interior collapse timescale (global collapse) is 0.32 for the sharp edged cylinder and only increases to 0.40 for the $\sigma = 1$ cylinder. It is not until the $\sigma = 4$ cylinder, in which the central, uniform density section of the cylinder is just 60\% of the length of the sharp edged cylinder, before the ratio of the minimum collapse timescale to the interior collapse timescale rises above 1/2. We therefore suggest that this edge collapse mode should still be significant in realistically tapered filaments. While we have not performed a similar calculation for a finite thickness, tapered disc, we expect that a very large taper would also be required before this edge effect becomes negligible in a disc.
			
\section{DISCUSSION}
\label{discussion}

\subsection{Interpretation of Analytical Results}
\label{interpretation}

	We have shown that weak density perturbations in spheres, discs, and filaments will collapse faster than the entire cloud, but that the significance of these local collapse modes depends strongly on the geometry of the cloud. 
	
	For spheres and discs, global collapse modes are significant. As shown in \S \ref{3d}, the timescale for the collapse of perturbations in spheres is only dependent upon the strength of the perturbation and thus, very large perturbations are required for a spherical cloud to fragment. The collapse timescale for perturbations in discs, as shown in \S \ref{2d}, is dependent upon the relative length scale of the perturbation, in comparison to the total size of the disc, but since thermal support sets a minimum size scale for collapse, roughly a Jeans length, we find that discs must be unrealistically large, having masses greater than 10$^4$ Jeans masses, for weak perturbations to collapse reasonably quicker than the entire disc. As such, discs also require large perturbations, possibly seeded by supersonic turbulence, in order for significant fragmentation to occur.
		
	In \S \ref{1d}, we find, however, that filamentary geometry provides a favorable situation for small perturbations to grow before global collapse overwhelms them. This situation occurs because the local collapse timescale in filaments has a stronger dependance on the length scale of the perturbation than in discs or spheres. A filament needs to be only a few Jeans lengths long in order for local collapse to occur much more rapidly than global collapse, even with only weak perturbations. Thus, once filaments are formed, they are likely to fragment into multiple objects, independent of the nature of turbulence in the filaments, which suggests that star forming cores should readily occur in filamentary structures in molecular clouds. Recent observations have discovered a correlation between star forming cores and dense filaments, as discussed further in \S \ref{obs}.
	
	We have only examined the collapse timescales along the longest axis of our clouds and have assumed that our clouds are supported from collapsing along their short axis. This is not unreasonable as pressure support is most effective on small scales (i.e., scales smaller than the Jeans length). For an infinite, isothermal disc, regardless of the surface density, there is always an equilibrium configuration of material along the short axis in which the disc has a thickness of approximately a Jeans length \citep{Spitzer42, Spitzer78} and for isothermal spheres, there is no shorter axis and thus, no extra collapse mode to account for. For an isothermal, infinite filament, there is a critical mass per unit length for the filament to be stable against radial collapse and at this critical mass per unit length, the filament has a radius of approximately a Jeans length \citep{Stodolkiewicz63, Ostriker64}. Below this critical value, a filament can be pressure confined (e.g., \citealt{Inutsuka97}), but above this critical value, radial collapse cannot be prevented by thermal support alone.
	
	Our long axis collapse modes increase the mass per unit length of a filament and could easily raise the mass per unit length of a subcritical filament above the critical value. If such a filament contains a weak perturbation, the perturbation will reach the critical mass per unit length before the rest of the filament and thus, radial collapse will occur first in the perturbation. This radial collapse will increase the speed at which the perturbation collapses and thus, will make fragmenting a filament even easier than what our analytic work suggests. 
	
	While the radial collapse mode may determine the late time density evolution of a perturbation, collapse along the long axis should still determine the final mass of the object as the majority of the mass accessible to a perturbation is along the long axis of the filament. Since our analytic results show that perturbations on the smallest length scales collapse the quickest and since thermal pressure sets a minimum collapse length scale at the Jeans length, filaments should fragment to form roughly Jeans mass objects separated by a few Jeans lengths.
	
	This analytic work assumes that there are no effective large scale support mechanisms operating within a molecular cloud. If such a mechanism were present, molecular clouds could survive for multiple free fall times and even slow local collapse modes could occur.  Possible support mechanisms discussed in the literature include driven turbulence \citep{Bonazzola92, Krumholz06, Li06} and magnetic fields \citep{Mouschovias76, Nakano76, Shu77, Shu87}. 
	
	\citet{BallesterosParedes09VazquezSemadeni} show that the galactic gravitational potential has a relatively negligible effect on a spherical molecular cloud, but for a filamentary cloud, the external potential can be larger than the self-gravity of the cloud. \citet{BallesterosParedes09Loinard} examine the energy budget of the Taurus molecular cloud and find that over the entire cloud, the galactic tidal energy is larger than the gravitational energy and should act to disrupt the cloud. On the smaller scales of clumps, however, \citet{BallesterosParedes09Loinard} find that the gravitational energy is much larger than the tidal energy and thus, in the Taurus molecular cloud, the galaxy's gravitational potential should prevent global collapse but should not significantly hinder small scale collapse.
	
\subsection{Comparison with Simulations and Observations}
\label{comparison}

In this section, we apply the results of our analysis to both simulations, in \S \ref{sims}, and observations, in \S \ref{obs}.

\subsubsection{Simulations}
\label{sims}

Simulations of supersonic turbulence show that the fragmentation of three-dimensional clouds into sheets and filaments is a natural consequence of supersonic turbulence (e.g., \citealt{Bate02, Bate03, Klessen00, Ostriker01, Clark04, Li04, VazquezSemadeni05}). Magnetic fields are also only capable of providing support perpendicular to the field direction and thus, magnetically supported clouds will also tend to collapse to form sheets and filaments. As we have shown, once filamentary substructure forms, the filaments can readily fragment to form star forming cores. Thus, star formation may occur through a process of reduction of dimensions, wherein spherical and disc shaped clouds first fragment to form filaments and then star formation occurs within these filamentary structures (see also \citealt{Larson85}).

Sheet-like, or at least flattened, clouds would be expected if clouds assemble in large-scale flows from the ambient medium. Besides observational evidence (see e.g., \citealt{Beaumont10}), a whole series of numerical experiments has demonstrated that the formation of clouds due to sweep-up by shells or flow collisions can lead to flattened (yet turbulent) structures \citep{VazquezSemadeni07, Heitsch08, Banerjee09} that, in some cases, display ring-like structures, as seen in the more idealized experiments by \citet{Burkert04}. \citet{Heitsch08Slyz} explicitly demonstrate that the ring-like filaments in their models are caused by global gravitational modes at the edge of the cloud. Yet from the current models it is difficult to see whether these ring structures will persist in a more general situation, where the colliding streams are not strongly spatially constrained. Unlike the idealized, truly two-dimensional models of \citet{Burkert04}, the three-dimensional cloud-formation models do show local collapse modes induced by turbulence and/or strong cooling (see \citealt{Heitsch08HartmannBurkert} for a discussion of the timescales). 

The preferential formation of density enhancements at the edges of filaments and discs, as would be caused by the edge collapse mode described in \S \ref{edge}, does occur in various numerical simulations (e.g., \citealt{Bastien83, Hartmann07}). In \S \ref{edge}, we determined that while this edge collapse mode can be suppressed in filamentary structures by tapering the edges of the filament, the taper has to be quite large in order for the edge collapse mode not to be significant. \citet{Li01} examines the importance of sharp edges to the formation of a dense ring along the outer edge of a collapsing disc and finds that the formation of this ring is only suppressed when the edge of the disc is tapered and the length scale of the taper is at least comparable to the size of the uniform density portion of the disc. Similarly, for filaments, the simulations of \citet{Nelson93} show that  prolate spheroids, which are highly tapered cylinders, do not always form fragments at their ends. 

Consideration of the short global free-fall time for finite molecular clouds suggests that it is hard to maintain a ``quiescent" star formation mode whereby individual Jeans mass objects fall out of a  ``quasi-static" cloud. Rather, our results emphasize that the bulk of the cloud also partakes in the gravitational collapse. While observations of molecular clouds often refer to ``quiescent" versus ``clustered" star formation, the distinction in this work is between local and global collapse in a finite cloud. In our picture, global collapse inevitably occurs, but local collapse is more prone to happen at lower dimensions, namely in rings/edges for sheets, and at the ends of filaments. Yet, note that this kind of local collapse occurring at the edges of clouds is still at least partially driven by global accelerations in all cases. Likewise, local, pre-seeded perturbations are more likely to grow and collapse in clouds of lower dimensionality; perturbations in sheets, if large enough, and filaments have more of a chance to collapse before global collapse ensues. 
		
	The similarity between the local and global collapse timescales in spheres and discs means that global collapse modes cannot be ignored in models or simulations of spherical or disc shaped molecular clouds. Thus, predictions of evolution timescales from models of cloud fragmentation in periodic boxes \citep{Klessen00, Ostriker01, Heitsch01, Li04, Padoan02, VazquezSemadeni05}, or quasi-static approaches \citep{Myers09}, are strongly limited by their assumptions of boundary conditions and/or initial states. Furthermore, since periodic box models do not incorporate physical processes occurring on scales larger than the box length, such as global collapse, it is exceedingly difficult to accurately run them for longer than the background global collapse timescale. As such, while periodic box simulations accurately follow the evolution of strong density perturbations, they cannot follow the collapse of small perturbations to completion. In other words, they only give meaningful results on timescales that are shorter than the global collapse timescale in the absence of the periodic boundaries. 

\subsubsection{Observations}
\label{obs}

Filamentary structure in star forming clouds appears to be ubiquitous. Filaments are present in both high mass and low mass star forming regions and are detected from the largest scales of molecular clouds all the way down to the scale of individual star forming cores; although, individual cores tend to be less elongated as thermal pressure plays a larger role on smaller spatial scales (e.g., \citealt{Schneider79, Scalo85, Bally87, Heyer87, Kulkarni88, Loren89, Nozawa91, Tatematsu93, Onishi96, Lada99, Jijina99, Heithausen02,  Hartmann02, Johnstone03, Lada07, Myers09, Molinari10, Andre10}). For instance, the Orion A molecular cloud is filamentary on large scales and contains filamentary substructure, including the integral shaped filament and the Orion Nebula Cluster (ONC) (e.g., \citealt{Bally87, Johnstone99}). Such filamentary structure is seen in both gas tracers and in star counts (e.g., \citealt{Bally87, Hillenbrand98}). Radial velocity measurements of stars and gas around the ONC have revealed a large scale velocity gradient, which has been interpreted as global collapse along the long axis of the filamentary structure of the ONC \citep{Furesz08, Tobin09}. The large scale structure of the Orion A molecular cloud has also been remarkably well reproduced by a simulation of a rotating disc with a density gradient \citep{Hartmann07}. 

The presence of filamentary structures appears to be closely tied to the star formation process, as young protostars and bound prestellar objects are preferentially located within dense filaments \citep{Hartmann02, Andre10}. \citet{Andre10} further note that Class 0 protostars and bound prestellar cores primarily exist in filaments that have mass per unit lengths larger than the critical mass per unit length at which an infinite, isothermal cylinder will be unstable to radial collapse (see \S \ref{thermal}), suggesting that the radial collapse mode of filaments plays a vital role in fragmenting a molecular cloud. This association of star forming cores with filaments is what we would expect given that our analytical results indicate that only filaments can readily fragment without strong perturbations. 

\section{CONCLUSIONS}
\label{conclusion}
	
	We have analytically estimated the global collapse timescales of spherical, disk shaped, and filamentary molecular clouds, as well as the local collapse timescales for small perturbations in these same clouds. We have shown that local collapse modes with timescales less than the timescales for global collapse do exist for perturbed spheres, discs, and cylinders and that local collapse is more effective for lower dimensional objects. We find the square of the ratio of global to local collapse timescales is inversely dependent upon the square of the relative size of the perturbed region of a cylinder, is inversely dependent upon the relative size of the perturbed region of a disc and is independent of the relative size of the perturbed region of a sphere. Table \ref{tab:analyticsummary} summarizes the most important derived equations relating to the collapse of perturbed spheres, discs, and cylinders. 
	
	We find that filamentary geometry is the most favorable situation for the smallest perturbations to grow before global collapse overwhelms them and that filaments containing only a few Jeans masses and weak density perturbations can readily fragment. Conversely, we find that weak perturbations in realistically sized discs and spherical clouds do not collapse significantly faster than the entire cloud.  Global collapse modes are significant in spheres and discs and cannot be ignored in simulations or models of such clouds. We also find that there exists an edge instability in discs and filaments that may lead to dense, locally collapsing structures along the periphery of these clouds. Our results suggest that star formation may occur through a process of reduction of dimensions, wherein spherical and disc shaped clouds first fragment to form filaments and then star formation occurs within these filamentary structures 
	
We would like to thank Lee Hartmann for originally bringing to our attention the tension between local and global collapse modes, as well as for many insightful discussions regarding the potential importance of global collapse modes. We would also like to thank Dr. Hartmann for a critical reading of an early version of this paper that has greatly improved the quality of this manuscript. We would also like to thank our anonymous referee for many useful changes to this paper. AP was partially supported by the Natural Sciences and Engineering Research Council of Canada graduate scholarship program. DJ acknowledges support from an NSERC Discovery Grant. FH gratefully acknowledges support by the NSF through grant AST 0807305, and by the NHSC through grant 1008. This research has made use of NASA's Astrophysics Data System.
	
\bibliographystyle{apj}
\bibliography{ponbib}{}

\newpage

\begin{landscape}
\begin{deluxetable}{cccc}
\tablecolumns{4}
\tablecaption{Summary of Key Analytic Equations \label{tab:analyticsummary}}
\tablewidth{0pt}
\tablehead{
\colhead{Quantity} & \colhead{3D Sphere} & \colhead{2D Disc} & \colhead{1D Cylinder}}
\startdata
$a_{exact}$ & $\frac{4\pi \,G \, \rho \, r}{3}\left(1 + \frac{\epsilon\,R_1^3}{r^3}\right)$ & $4G \, \Sigma \frac{R}{r} \Bigg( \Big[K\big(\frac{r} {R}\big)   - E\big(\frac{r}{R}  \big)\Big] + \epsilon \frac{r}{R}\, \Big[K\big(\frac{R_1} {r}\big) - E\big(\frac{R_1}{r}\big)\Big]\Bigg)$ & $\frac{G \lambda}{L - d} - \frac{G \lambda} {L + d} + \frac{G \lambda \, \epsilon} {d - L_1} - \frac{G \lambda \epsilon} {L_1 + d}$ \\

$a_{approximate}$ & $\frac{4\pi \,G \, \rho \, r}{3}\left(1 + \frac{\epsilon\,R_1^3}{r^3}\right)$ & $\frac{\pi G \, \Sigma \, r}{R} + \frac{\pi G \, \epsilon \Sigma \, R_1^2} {r^2}$ & $\frac{G \lambda \, 2d}{L^2} \left(1+\frac{\epsilon\,L_1\,L^2}{d^3}\right)$ \\

$t^2$ & $\frac{3}{2\pi \,G \, \rho} \left(1 + \frac{\epsilon\,R_1^3}{r^3}\right)^{-1}$ & $\frac{2R}{\pi G \, \Sigma}\left(1 + \frac{\epsilon \, R \, R_1^2}{r^3}\right)^{-1}$ & $\frac{L^2}{G \lambda} \left(1+\frac{\epsilon\,L_1\,L^2}{d^3}\right)^{-1}$ \\

$\frac{t^2_g}{t^2_l}$ &  $1+ \epsilon$ & $1+ \frac{\epsilon \, R}{R_1}$ & $1+ \frac{\epsilon \, L^2}{L_1^2}$ \\ 
\enddata
\end{deluxetable}
\end{landscape}

\end{document}